\documentclass[12pt]{article}
\usepackage{amsfonts}
\usepackage{graphicx}
\usepackage{amsmath}
\usepackage{float}
\textwidth=6.5in \hoffset=-.75in \voffset=-1in \numberwithin{equation}{section}
\numberwithin{figure}{section}

\newcommand {\nn}{\nonumber}
\newcommand {\be}{\begin{equation}}
\newcommand {\ee}{\end{equation}}
\begin{document}

\begin{titlepage}
\vspace{1cm}
\begin{center}
{\Large \bf {Exact convoluted solutions in higher-dimensional Einstein-Maxwell theory}}\\
\end{center}
\vspace{2cm}
\begin{center}
A. M. Ghezelbash{ \footnote{ E-Mail: masoud.ghezelbash@usask.ca}}
\\
Department of Physics and Engineering Physics, \\ University of Saskatchewan, \\
Saskatoon, Saskatchewan S7N 5E2, Canada\\
\vspace{1cm}
\vspace{2cm}
%\today\\
\end{center}

\begin{abstract}
We construct two classes of exact solutions to six and higher dimensional Einstein-Maxwell theory in which the metric functions can be written as convolution-like integrals of two special functions. The solutions are regular everywhere and show a bolt structure on a single point in any dimensionality.
Moreover, we find the exact nonstationary solutions to the Einstein-Maxwell theory with positive cosmological constant. We show that the cosmological solutions are expanding patches in asymptotically de Sitter spacetime.

\end{abstract}
\end{titlepage}\onecolumn 
\bigskip 

\section{Introduction}
The exact solutions to Einstein-Maxwell gravity are of great significance in exploring the different aspects of gravitational physics. More specifically, exploring the higher dimensional solutions to Einstein-Maxwell theory (and its generalization which include dilaton or axion field as well as Chern-Simons term) provide a treasure trove of possibilities that normally do not exist in four dimensions. The existence of black holes with 3-sphere horizon \cite{Myers1}, black rings with $S^2 \times S^1$ horizon \cite{Em1}, black lens \cite{Ch1} and black saturn \cite{El1} in five dimensions are just few interesting examples. One can find solutions to higher dimensional Einstein-Maxwell equations which are asymptotically locally flat (dS or AdS) with non-vanishing NUT charges \cite{awad}. The other interesting solutions include the solitonic and dyonic solutions \cite{EY2,EY22}, braneworld cosmologies \cite{EY4} and string theory extended solutions \cite{EY5}. The supegravity solutions and cosmological solutions and solutions with dilaton, axion and Chern-Simons term are considered in \cite{other13}-\cite{other16}.
Moreover, in an interesting paper \cite{hashi}, the authors constructed the convolution-like solutions for the fully localized type IIA
D2 branes intersecting D6 branes. The solutions are obtained by compactifying the corresponding convolution-like M2 brane solutions 
over a circle of transverse self-dual Taub-NUT geometry. The solutions preserve eight supersymmetries and are valid everywhere; near and far the core of D6 branes. 
One feature of the solutions is that the solutions are expressed completely in terms of convoluted integrals that is a result of taking special ansatz for the solutions and separability of the equations of motion \cite{me}.
According to our knowledge, there are not any known convoluted solutions in six dimensional or higher dimensional Einstein-Maxwell theories. Inspired by the convoluted M2-brane solutions, in this article, we construct such convoluted solutions in six and higher dimensional Einstein-Maxwell theory. To have non-trivial convoluted solutions, the minimal dimensionality of Einstein-Maxwell theory should be six. Moreover, we consider the Einstein-Maxwell theories with positive cosmological constant in six and higher dimensions and find the exact cosmological convoluted solutions. The outline of the paper is as follows. In section \ref{sec:6D}
we present the convoluted solutions for the six-dimensional Einstein-Maxwell theory by superimposing the solutions to the equations of motion. We then fix the measure functions in the convoluted integral by comparing the appropriate limits of the solutions to that of exact flat solutions to the Einstein-Maxwell theory. Moreover, we present the convoluted solutions for the seven dimensional theory. Inspired by the solutions found in section \ref{sec:6D}, we construct the exact convoluted solutions for the $D$-dimensional Einstein-Maxwell theory in section \ref{sec:DD}.
In section \ref{sec:2ndsolutions}, we find the second set of solutions in $D$-dimensions by analytically continuing the separation constant.
In section \ref{sec:cosmo}, we consider the Einstein-Maxwell theory with positive cosmological constant and show the equations of motion are separable if one consider a proper separation of three coordinates in the metric functions. The concluding remarks wrap up the paper in \ref{sec:con}.

\section{Convoluted solutions in six and seven-dimensional Einstein-Maxwell theory }
\label{sec:6D}
\begin{figure}[H]
\centering
\includegraphics[width=0.5\textwidth]{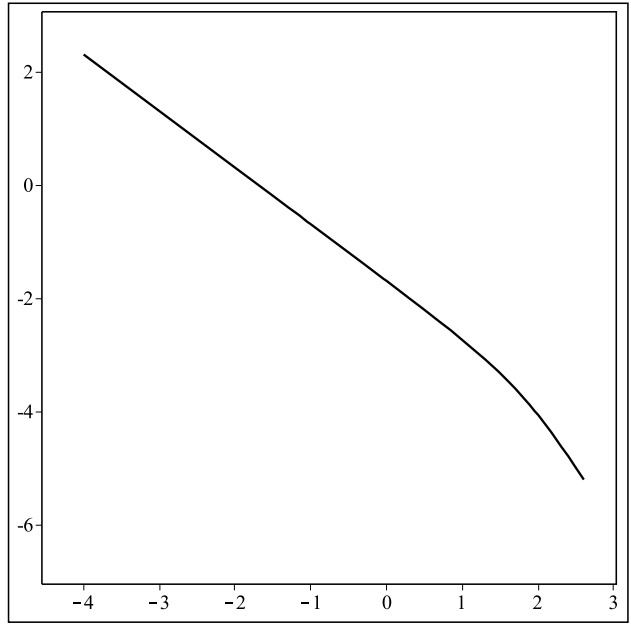}
\caption{The logarithmic plot of general solution (\ref{HTN}) ($\log(H(r,0)-1)$ versus $\log(r)$)}
\label{fig:figure1}
\end{figure}

We consider the six-dimensional line element
\begin{equation}
ds_6^{2}=-H(r,x)^{-2}dt^{2}+H(r,x)^{2/3}(dx^2+ds_{n}^2),
\label{ds6}
\end{equation}
where the four-dimensional space $ds_{n}^2$ with positive NUT charge $n$ is given by
\begin{equation}
ds_{n}^{2}=V(r)(dr^{2}+r^2d\Omega^2) +\frac{(d\psi+n\cos\theta d\phi)^2}{V(r)}.\label{Nmetric}
\end{equation}%
The coordinate $\psi$ belongs to interval $[0,4\pi n]$ which parameterizes the fibration of a circle over the sphere and $r \geq 0$ and the $r$-dependent metric function $V(r)$ is $V(r)=1+\frac{n}{r}$. We take the only non-vanishing component of the gauge field as
\be
{A_t}=\frac{2}{\sqrt{3}H(r,x)}\label{gauge}.
\ee
We note that the metric (\ref{Nmetric}) can also be written as 
\be
ds_{n}^2=\frac{1}{f(r)}(d\psi +n \cos \theta d\phi)^2+f(r)dr^2+(r^2-\frac{n^2}{4})(d\theta^2+\sin ^2\theta d\phi^2)\label{TN},
\ee
by a shift in the radial coordinate. The radial coordinate in (\ref{TN}) is greater or equal to $\frac{n}{2}$ and the
radial function $f(r)$ is given by
\be
f(r)=\frac{2r+n}{2r-n}.
\ee
The gravitational and electromagnetic field equations are respectively given by
\begin{eqnarray}
R_{\mu\nu}&=&F_{\mu}^{\lambda}F_{\nu\lambda}-\frac{1}{8}F^2,\label{eq1}\\
F^{\mu\nu}_{;\mu}&=&0,\label{eq2}
\end{eqnarray}
that imply the metric function $H(r,x)$ satisfies the second order partial differential equation
\be
V \left( r \right) \left( {\frac {\partial ^{2}}{\partial {x}^{2}}}H
\left( r,x \right) \right) r+ \left( {\frac {\partial ^{2}}{
\partial {r}^{2}}}H \left( r,x \right) \right) r+2\,{\frac {\partial 
}{\partial r}}H \left( r,x \right)=0.\label{Feq}
\ee
To solve the differential equation (\ref{Feq}), we separate the coordinates as $H(r,x)=1+hH_1(r)H_2(x)$ that leads to two ordinary differential equations for $H_1(r)$ and $H_2(x)$
\be
{\frac {{\frac {d^{2}}{d{r}^{2}}}{\it H_1} \left( r \right) }{{\it H_1}
\left( r \right) V \left( r \right) }}+2\,{\frac {{\frac {d}{dr}}{
\it H_1} \left( r \right) }{{\it H_1} \left( r \right) rV \left( r
\right) }}-\epsilon\,C^2=0,\label{H1}
\ee
and
\be
{\frac {{\frac {d^{2}}{d{x}^{2}}}{\it H_2} \left( x \right) }{{\it H_2}
\left( x \right) }}+\epsilon\,C^2=0,\label{H2}
\ee
where $C$ is the separation constants and $\epsilon=\pm 1$. The solutions to differential equation (\ref{H1}) are 
\be
H_1(r)={h_{1M}}\,{{\rm e}^{-\sqrt {\epsilon}{C}r}}
{{\cal M}\left(1+1/2\,\sqrt {\epsilon} {C}n,\,2,\,2\,\sqrt {\epsilon}Cr\right)}
+{h_{1U}}\,{{\rm e}^{-\sqrt {\epsilon} {C}r}}
{{\cal U}\left(1+1/2\,\sqrt {\epsilon}{C}n,\,2,\,2\,\sqrt {\epsilon}{C}r\right)},\label{H1r}
\ee
in terms of Kummer ${\cal M}$ and ${\cal U}$ functions and $h_{1M}$ and $h_{1U}$ are two arbitrary constants. The equation (\ref{H2}) has the solutions
\be
H_2(x)=h_{2}\cos \left( \sqrt {\epsilon} {C}x+\beta \right),\label{H2x}
\ee
with $h_2$ and $\beta$ as constants. To construct the most general solution for the differential equation (\ref{Feq}), we consider the superposition of solutions (\ref{H1r}) and (\ref{H2x}) as
\be
H(r,x)=1+ \int _{0}^\infty dC\, f(C) \, {{\rm e}^{- {C}r}}
{{\cal U}\left(1+1/2\, {C}n,\,2,\,2\, {C}r\right)} \cos ({C}x+\beta),\label{HTNU}
\ee
where we choose $\epsilon=1,h_{1M}=0$ and $f(C)$ is any arbitrary measure function. To find the proper measure function, we notice that in the limits $r \rightarrow 0$ and $n\rightarrow \infty$, the metric (\ref{Nmetric}) describes a closed 4-ball $B^4$ with the line element $dw^2+w^2d\Omega_3^2$ where $w^2=4nr$. In this limiting case, we can find that the six-dimensional metric 
\be
ds^2=-\frac{1}{H_0^2(r,x)}dt^2+H_0(r,x)^{2/3}(dx^2+dw^2+w^2d\Omega_3^2),\label{metr6flat}
\ee
and the gauge field (\ref{gauge}) are the exact solutions to the equations of motion where the metric function $H_0(r,x)$ is 
\be
H_0(r,x)=1+\frac{\alpha}{(4nr+x^2)^{3/2}}\label{H06},
\ee
and $\alpha$ is a constant. Demanding the metric function (\ref{HTNU}) matches to (\ref{H06}) in the limits $r \rightarrow 0$ and $n\rightarrow \infty$, we find an integral equation for the measure function $f(C)$, given by
\be
\lim_ {r\rightarrow 0,n \rightarrow \infty}\int _0^\infty dC f(C)
\,{{\rm e}^{- {C}r}}
{{\cal U}\left(1+1/2\, {C}n,\,2,\,2\, {C}r\right)}\cos ({C}x+\beta)=\frac{\alpha}{(4nr+x^2)^{3/2}}.\label{inteq}
\ee
We can find the solutions to integral equation (\ref{inteq}), noting the identity 
\be
\lim _{r\rightarrow 0, n\rightarrow \infty}{{\cal U}\left(1+1/2\, {C}n,\,2,\,2\, {C}r\right)}=\frac{2}{C\sqrt{rn}\Gamma(Cn/2)}K_1(2C\sqrt{nr}),
\ee
where $K_1$ is the modified Bessel function. Hence the integral equation (\ref{inteq}) reduces to
\be
\int _0^\infty dC f(C)
\,
\frac{2}{C\sqrt{rn}\Gamma(Cn/2)}K_1(2C\sqrt{nr})\cos ({C}x+\beta)=\frac{\alpha}{(4nr+x^2)^{3/2}}.\label{inteq2}
\ee
After a lengthy calculation, we find that the only solution to the integral equation (\ref{inteq2}) is given by
\be f(C)=\frac{\alpha C^2\Gamma(Cn/2)}{2\pi},
\ee
along with the choice of $\beta=0$.
So finally, we find the exact convoluted solution for the metric function $H(r,x)$ as
\be
H(r,x)=1+ \frac{\alpha}{2\pi} \int _{0}^\infty dC\, C^2\Gamma(Cn/2) \, {{\rm e}^{- {C}r}}
{{\cal U}\left(1+1/2\, {C}n,\,2,\,2\, {C}r\right)} \cos ({C}x) \label{HTN}.
\ee
Although we can't find an analytic expression for the integral in equation (\ref{HTN}), however we can numerically calculate the integral and plot it in terms of coordinate $r$. Figure \ref{fig:figure1} shows the behaviour of function $\log(H(r,x=0)-1)$ versus $\log(r)$ where we set $\alpha=2\pi, n=1$. As we notice, the metric function is regular everywhere except around the location of bolt at $r=0$ in four-dimensional metric (\ref{Nmetric}). The Kretchmann invariant of the solution is finite at $x=0$, however near the location of bolt in four-dimensional metric (\ref{Nmetric}), the Kretchmann invariant of metric (\ref{ds6}) behaves as $K=\frac{k(x)}{r^{2/3}}$ where $k(x)$ is a regular function of $x$. 
We note that taking more coordinate dependence for the metric function $H$ may shield the point $r=0$ from physical region of interest, so making the solution regular everywhere. 
We consider now $\epsilon=1,h_{1U}=0$ that may lead to the second general solution. The superposition of solutions read as
\be
H(r,x)=1+ \int _{0}^\infty dC\, g(C) \, {{\rm e}^{- {C}r}}
{{\cal M}\left(1+1/2\, {C}n,\,2,\,2\, {C}r\right)} \cos ({C}x+\gamma), \label{HTNM}
\ee
where $g(C)$ is the measure function and $\gamma$ is a constant. To find the measure function $g(C)$, we compare the metric function (\ref{HTNM})
to (\ref{metr6flat}) in the limits where $r \rightarrow 0$ and $n\rightarrow \infty$ that leads to an integral equation for $g(C)$. In these limits, we find the identity
\be
\lim _{r\rightarrow 0, n\rightarrow \infty}{{\rm e}^{- {C}r}}{{\cal M}\left(1+1/2\, {C}n,\,2,\,2\, {C}r\right)}=\frac{1}{C\sqrt{rn}}I_1(2C\sqrt{nr}),
\ee
where $I_1$ is the modified Bessel function. Although we can find the solution to the integral equation for $g(C)$ that is given by $g(C)=\frac{\alpha C^2}{2}$ with $\gamma=0$, however the solution is not well defined where $r\rightarrow \infty$. In fact for $r\rightarrow \infty$,
the Kummer ${\cal M}$ function behaves as 
${{\cal M}\left(1+1/2\, {C}n,\,2,\,2\, {C}r\right)} \rightarrow e^{2Cr} (2Cr)^{-1+Cn/2}\{1+O(\frac{1}{2Cr})\}$, so the integrand in the metric function $H(r,x)$ is exponentially divergent at large values of $r$. We note that the solution (\ref{HTNU}) is well defined at large $r$ where the Whittaker ${\cal U}$ function approaches ${{\cal U}\left(1+1/2\, {C}n,\,2,\,2\, {C}r\right)} \rightarrow (2Cr)^{1+1/2Cn}\{1+O(\frac{1}{2cr})\}$. Furnished with the six dimensional solution (\ref{HTN}) and (\ref{ds6}), we consider the embedding of the metric (\ref{Nmetric}) into seven dimensional Einstein-Maxwell theory. For $D=7$, we consider the metric ansatz 
\be
ds_7^{2}=-H(r,x)^{-2}dt^{2}+H(r,x)^{1/2}(dx^2+x^2d\chi^2+ds_{n}^2),
\ee
where $0\leq \chi \leq 2\pi$ and the only non-zero component of the gauge field as 
\be
{A_t}=\frac{\sqrt{5}}{{2}H(r,x)}\label{gauge7}.
\ee
After separating the equations of motion by substituting $H(r,x)=1+hH_1(r)H_2(x)$, we find two ordinary differential equations for $H_1(r)$ and $H_2(x)$. The differential equation for $H_1(r)$ is the same as equation(\ref{H1}) and the solutions to differential equation for $H_2(x)$ are given by
\be
{\it H}_2 \left( x \right) ={\it h_{2J}}\,
{{J_0}\left(C x\right)}+{\it h_{2Y}}\,
{{Y_0}\left(C x\right)}.
\ee
So, we have the most general solution for the metric function $H(r,x)$ as
\be
H(r,x)=1+ \int _{0}^\infty dC\, {{\rm e}^{- {C}r}}
{{\cal U}\left(1+1/2\, {C}n,\,2,\,2\, {C}r\right)} \{f_J(C){{ J_0}\left(C x\right)}+f_Y(C){{Y_0}\left(C x\right)}\}\label{HTN77}
\ee
where $f_J(C)$ and $f_Y(C)$ are two arbitrary measure functions. To find the proper measure functions, as we noticed before, in the limits $r \rightarrow 0$ and $n\rightarrow \infty$, the transverse space describes $B^2\times B^4$. In this limiting case, we can find the exact solution for the seven-dimensional flat metric that is given by
\be
ds^2=-\frac{1}{H_0^2(r,x)}dt^2+H_0(r,x)^{1/2}(dx^2+x^2d\chi^2+dw^2+w^2d\Omega_3^2),
\ee
where $H_0(r,x)$ is
\be
H_0(r,x)=1+{\frac {\gamma}{ \left( {r}^{2}+{x}^{2} \right) ^{2}}}\label{H07}.
\ee
and $\alpha$ is a constant. Matching the metric function (\ref{HTN77}) to (\ref{H07}) in the limits $r \rightarrow 0$ and $n\rightarrow \infty$, leads to an integral equation for $f_J(C)$ and $f_Y(C)$. We find the unique solutions to the integral equation are $f_J(C)=\frac{C^3\gamma}{8}\Gamma(\frac{Cn}{2})$ and $f_Y(C)=0$. So, the most general solution for $H(r,x)$ is 
\be
H(r,x)=1+ \frac{\gamma}{2\pi} \int _{0}^\infty dC\, C^3\Gamma(Cn/2) \, {{\rm e}^{- {C}r}}
{{\cal U}\left(1+1/2\, {C}n,\,2,\,2\, {C}r\right)} J_0 ({C}x) \label{HTN7}.
\ee
\section{Convoluted solutions in $D\geq 6$ Einstein-Maxwell theory}
\label{sec:DD}
Furnished with the general convoluted solutions for the metric function $H(r,x)$ in six and seven dimensional Einstein-Maxwell theory, we try to find the general solutions for $H(r,x)$ in $D$ dimensional theory. We consider the $D$-dimensional metric ansatz as
\be
ds_{D}^{2}=-H(r,x)^{-2}dt^{2}+H(r,x)^{2/(D-3)}(dx^2+x^2d\Omega_{D-6}+ds_{n}^2),\label{mD}
\ee
where $d\Omega_{D-6}$ is the metric on unit sphere $S^{D-6}$. Moreover, we consider the only non-zero component of the gauge field as
\be
{A_t}={\sqrt{\frac{D-2}{D-3}}H(r,x)}\label{gaugeD}.
\ee 
Similar to six and seven dimensional cases, the equations of motion are separable if we consider $H(r,x)=1+hH_1(r)H_2(x)$. The differential equation for $H_1(r)$ is the same as (\ref{H1}) and the differential equation for $H_2(x)$ is 
\be
x{\frac {d^{2}}{d{x}^{2}}}{\it H_2} \left( x \right) + \left( D-6
\right) {\frac {d}{dx}}{\it H_2} \left( x \right) +\epsilon {\it H_2} \left( x
\right) x{C}^{2}=0. \label{H2Deq}
\ee
The solutions to (\ref{H2Deq}) are
\be
{\it H_2} \left( x \right) ={h_J}\,{x}^{\frac{7-D}{2}}
{{J_{\frac{D-7}{2}}}\left(Cx\right)}+{h_Y}\,{x}^{\frac{7-D}{2}}
{{Y_{\frac{D-7}{2}}}\left(Cx\right)},
\ee
and so, the general solution looks like
\begin{eqnarray}
H(r,x)&=&1+ \int _{0}^\infty dC\, {{\rm e}^{- {C}r}}
{{\cal U}\left(1+1/2\, {C}n,\,2,\,2\, {C}r\right)}\{{f_{J,D}(C)}\,{x}^{\frac{7-D}{2}}
{{J_{\frac{D-7}{2}}}\left(Cx\right)}\nonumber\\
&+&{f_{Y,D}(C)}\,{x}^{\frac{7-D}{2}}
{{Y_{\frac{D-7}{2}}}\left(Cx\right)}\} \label{HTN9},
\end{eqnarray}
where $f_{J,D}(c)$ and $f_{Y,D}(C)$ are two arbitrary measure functions. As we notice, in the limits $r \rightarrow 0$ and $n\rightarrow \infty$, the transverse space is $B^{D-5}\times B^4$. In this limiting case, we find the following exact solution for the $D$-dimensional metric 
\be
ds^2=-\frac{1}{H_{0,D}^2(r,x)}dt^2+H_{0,D}(r,x)^{2/(D-3)}(dx^2+x^2d\Omega_{D-6}+dw^2+w^2d\Omega_3^2),\label{nonut}
\ee
where the metric function $H_0(r,x)$ is
\be
H_{0,D}(r,x)=1+{\frac {B}{ \left( {r}^{2}+{x}^{2} \right) ^{\frac{D-3}{2}}}}\label{H09},
\ee
and the gauge field is still given by (\ref{gaugeD}). In (\ref{H09}), $B$ is a constant. We then find the following integral equation that we solve to determine and fix the measure functions $f_{J,D}(c)$ and $f_{Y,D}(C)$
\begin{equation}
\int _{0}^\infty dC\, {{\rm e}^{- {C}r}}
{{\cal U}\left(1+1/2\, {C}n,\,2,\,2\, {C}r\right)}{x}^{\frac{7-D}{2}}\{{f_{J,D}(C)}\,
{{J_{\frac{D-7}{2}}}\left(Cx\right)}
+{f_{Y,D}(C)}\,
{{Y_{\frac{D-7}{2}}}\left(Cx\right)}\}={\frac {B}{ \left( {r}^{2}+{x}^{2} \right) ^{\frac{D-3}{2}}}}\label{inteqD}.
\end{equation}
We find the solutions to the integral equation (\ref{inteqD}) are given by
$f_{Y,D}(C)=0$ for $D\geq 6$ and
\be
f_{J,D}(C)=B\frac{C^{\frac{D-1}{2}}}{\phi_D}\Gamma(\frac{Cn}{2}).
\ee
The $\phi_D$ for even dimensions $D=2k,\, k=3,4,\cdots$ is equal to 
\be
\phi_{2k}=2\sqrt{2\pi}\prod_{n=0}^{k-3}(2n+1)\label{phieven},
\ee
while for odd dimensions $D=2k+1,\, k=3,4,\cdots$ is 
\be
\phi_{2k+1}=4\prod_{n=1}^{k-2}(2n)\label{phiodd}.
\ee
Hence we find the most general $D$-dimensional solution as
\be
H(r,x)=1+ \frac{B}{\phi_D} {x}^{\frac{7-D}{2}}\int _{0}^\infty dC\, C^{\frac{D-1}{2}}\Gamma(Cn/2) \, {{\rm e}^{- {C}r}}
{{\cal U}\left(1+1/2\, {C}n,\,2,\,2\, {C}r\right)} 
{{J_{\frac{D-7}{2}}}\left(Cx\right)} \label{HTN7}.
\ee
\section{The second class of $D$-dimensional solutions}
\label{sec:2ndsolutions}

\begin{figure}[H]
\centering
\includegraphics[width=0.5\textwidth]{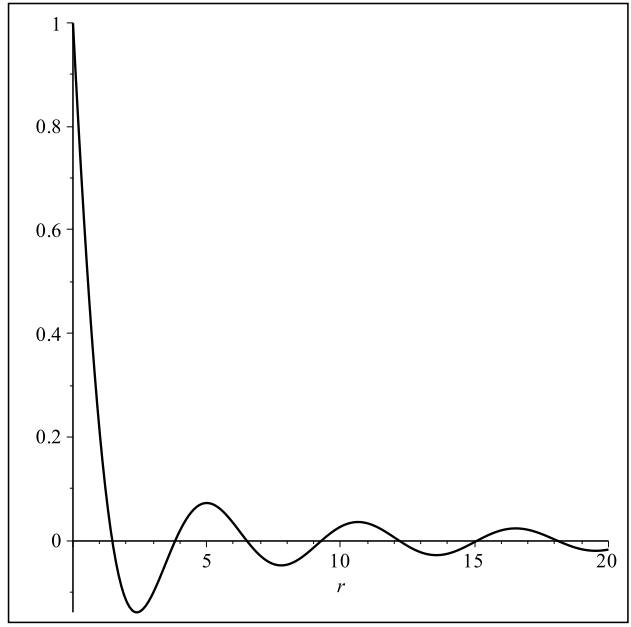}
\caption{The real and finite second solution for $H_1(r)$}
\label{fig:figure2}
\end{figure}

In all solutions presented in sections (\ref{sec:6D}) and (\ref{sec:DD}), we considered $\epsilon=1$. However, we can find a second set of solutions by taking $\epsilon=-1$. In this case, the solutions to equation (\ref{H1}) for $H_1(r)$ read as 
\be
H_1(r)={h'_{1M}}\,{{\rm e}^{-i{C}r}}
{{\cal M}\left(1+1/2\,i {C}n,\,2,\,2\,iCr\right)}
+{h'_{1U}}\,{{\rm e}^{-i {C}r}}
{{\cal U}\left(1+1/2\,i{C}n,\,2,\,2\,i{C}r\right)},
\ee
where ${h'_{1M}}$ and ${h'_{1U}}$ are two constants. To have a real valued function $H_1(r)$, we choose $h'_{1U}=0$. The resulting solution then is real and finite everywhere. In fact, the function ${{\rm e}^{-i {C}r}}
{{\cal M}\left(1+1/2\,i{C}n,\,2,\,2\,i{C}r\right)}$ is finite with value of 1 at $r=0$ and oscillating to zero at far infinity. Figure \ref{fig:figure2} shows the behaviour of this function where we set $n=2,C=1$. The real solution for $H_2(x)$ is given by $\cosh(Cx)$.
Superimposing all the solutions with different value of separation constant $C$, we find the second general solution for the six-dimensional Einstein-Maxwell theory 
\be
H(r,x)=1+ \int _{0}^\infty dC \, {{\rm e}^{-i {C}r}}
{{\cal M}\left(1+1/2\, i{C}n,\,2,\,2\, i{C}r\right)} \cosh ({C}x) \label{HTNM26}.
\ee
We can find the second general solutions in $D$-dimensions, noting that the solutions to equation (\ref{H2Deq}) with $\epsilon=-1$ for $H_2(x)$ are 
\be
{\it H_2} \left( x \right) ={h'_{2I}}\,{x}^{\frac{7-D}{2}}
{{ I}_{\frac{D-7}{2}}\left(Cx\right)}+{h'_{2K}}\,{x}^{\frac{7-D}{2}}
{{ K}_{\frac{D-7}{2}}\left(Cx\right)},
\ee
in terms of modified Bessel $I$ and $K$ functions. So, we find the second general convoluted solution to the $D$-dimensional Einstein-Maxwell theory as
\begin{eqnarray}
H(r,x)&=&1+ \int _{0}^\infty dC\, {{\rm e}^{-i {C}r}}
{{\cal M}\left(1+1/2\, i{C}n,\,2,\,2\, i{C}r\right)}\{{g_{I,D}(C)}\,{x}^{\frac{7-D}{2}}
{{ I_{\frac{D-7}{2}}}\left(Cx\right)}\nonumber\\&+&{g_{K,D}(C)}\,{x}^{\frac{7-D}{2}}
{{ K_{\frac{D-7}{2}}}\left(Cx\right)}\}, \label{HTNM2D}
\end{eqnarray}
where ${g_{I,D}(C)}$ and ${g_{K,D}(C)}$ are the measure functions. Comparing the general solution (\ref{HTNM2D}) to (\ref{H09}) in the proper limits and solving the integral equation, we find that the measure function 
$g_{I,D}(C)=0$ for $D\geq 6$ and the other measure function is
\be
g_{K,D}(C)=2B{C^{\frac{D-1}{2}}}/{\phi_D}.
\ee
We note that $\phi_D$ is given by (\ref{phieven}) and (\ref{phiodd}) for even and odd dimensions $D$, respectively. So, we find the second convoluted metric function in $D$-dimensions as
\begin{equation}
H(r,x)=1+ 2\frac{B}{\phi_D}{x}^{\frac{7-D}{2}}\int _{0}^\infty dC\, C^{\frac{D-1}{2}}\, {{\rm e}^{-i {C}r}}
{{\cal M}\left(1+1/2\, i{C}n,\,2,\,2\, i{C}r\right)}
{{ K_{\frac{D-7}{2}}}\left(Cx\right)}, \label{HTNS}
\end{equation}

\section{The cosmological convoluted solutions in $D\geq 6$ dimensional Einstein-Maxwell Theory}
\label{sec:cosmo}
\begin{figure}[H]
\centering
\includegraphics[width=0.5\textwidth]{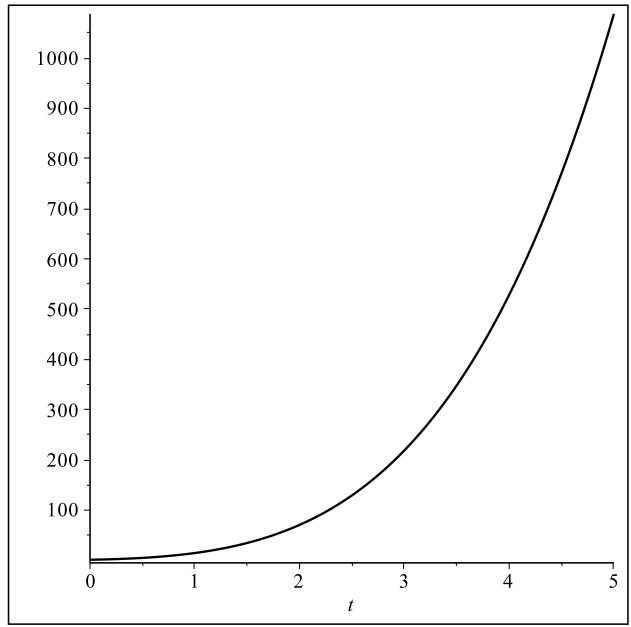}
\caption{The $c$ function versus time for a $r$-constant slice on six dimensional cosmological spacetime with the metric function (\ref{HTN6cos})}
\label{fig:c}
\end{figure}

In this section, we consider the Einstein-Maxwell theory in presence of positive cosmological constant. We first consider $D=6$ and take
the metric ansatz as 
\be
ds_6^{2}=-H(t,r,x)^{-2}dt^{2}+H(t,r,x)^{2/3}(dx^2+ds_{n}^2),\label{D6Cos}
\ee
where the metric function depends on $t$ as well as the spatial directions $r,x$.
The equations of motion in presence of cosmological constant lead to 
the following second order partial differential equation for $H(t,r,x)$
\begin{eqnarray}
&&-2H^{11/3}V^3r^4\frac{\partial ^2 H}{\partial t^2}-\frac{10}{3}H^{8/3}V^3r^4(\frac{\partial H}{\partial t})^2+2V^3r^4H\frac{\partial ^2 H}{\partial x^2}-2V^3r^4(\frac{\partial H}{\partial x})^2-3VH^2r^4\frac{\partial ^2 V}{\partial r^2}\nn\\
&+&2V^2r^4H\frac{\partial ^2 H}{\partial r^2}-2V^2r^4(\frac{\partial H}{\partial r})^2+3H^2r^4(\frac{\partial V}{\partial r})^2+4V^2H
r^3\frac{\partial H}{\partial r}-6VH^2r^3\frac{\partial V}{\partial r}-3n^2H^2\nn\\
&+&3\Lambda H^{8/3}V^3r^4+2V^3r^4(\frac{\partial H}{\partial x})^2+2V^2r^4(\frac{\partial H}{\partial r})^2=0.
\label{Geq}
\end{eqnarray} 
We can solve the partial differential equation\label{Geq} by separating the coordinates as $H(t,r,x)=H_1(r)H_2(x)+H_3(t)$. Quite interestingly we find that $H_1(r)$ and $H_2(x)$ satisfy the equations (\ref{H1}) and (\ref{H2}), while the solutions for $H_3(t)$ are given by $H_3(t)=h_3 t+h_4$ where $h_3=3\sqrt{\frac{\Lambda}{10}}$. To get agreement with the metric function (\ref{HTN}) in asymptotically flat spacetimes where $\Lambda=0$, we choose $h_4=1$. 
Moreover in presence of cosmological constant, we find the exact solution 
\be
ds^2=-\frac{1}{H_0^2(t,r,x)}dt^2+H_0(t,r,x)^{2/3}(dx^2+dw^2+w^2d\Omega_3^2),
\ee
to the Einstein-Maxwell theory along with the gauge field (\ref{gaugeD}) where the metric function $H_0(r,x)$ is given by
\be
H_0(t,r,x)=1+h_3t+\frac{\alpha}{(4nr+x^2)^{3/2}}\label{H0},
\ee
and $\alpha$ is a constant. So, the most general cosmological solution in six dimensions read as 
\be
H(t,r,x)=1+\sqrt{\frac{9\Lambda}{10}} t+ \frac{\alpha}{2\pi} \int _{0}^\infty dC\, C^2\Gamma(Cn/2) \, {{\rm e}^{- {C}r}}
{{\cal U}\left(1+1/2\, {C}n,\,2,\,2\, {C}r\right)} \cos ({C}x) \label{HTN6cos},
\ee
after matching with (\ref{H0}) in the appropriate limits and fixing the measure integral.
Similar calculations show in $D$ dimensions, we get two general cosmological convoluted solutions, given by
\be
H(t,r,x)=1+\sqrt{{\xi_D\Lambda}} t+ \frac{B}{\phi_D} {x}^{\frac{7-D}{2}}\int _{0}^\infty dC\, C^{\frac{D-1}{2}}\Gamma(Cn/2) \, {{\rm e}^{- {C}r}}
{{\cal U}\left(1+1/2\, {C}n,\,2,\,2\, {C}r\right)} 
{{J_{\frac{D-7}{2}}}\left(Cx\right)} \label{HTNDcos},
\ee
and
\be
H(r,x)=1+\sqrt{{\xi_D\Lambda}} t +\frac{2B}{\phi_D} {x}^{\frac{7-D}{2}}\int _{0}^\infty dC\, C^{\frac{D-1}{2}}\, {{\rm e}^{-i {C}r}}
{{\cal M}\left(1+1/2\, i{C}n,\,2,\,2\, i{C}r\right)} 
{{K_{\frac{D-7}{2}}}\left(Cx\right)} \label{HTNDsecondcos},
\ee
where
\begin{equation}
\xi_{D}=\frac{2(D-3)^2}{(D-1)(D-2)}.
\end{equation}
Writing the cosmological constant in terms of cosmological horizon ${\ell }$ as $\Lambda=\frac{(D-1)(D-2)}{2}$, we find the time dependent term in (\ref{HTNDcos}) and (\ref{HTNDsecondcos}) is simply $\sqrt{{\xi_D\Lambda}}t=\frac{D-3}{\ell}t$.
Since the general solutions (\ref{HTNDcos}) and (\ref{HTNDsecondcos}) describe the asymptotically dS spacetime, we 
take a look at the cosmological $c$-function for the solutions in context of dS/CFT correspondence. For any asymptotically $D$-dimensional dS spacetime, one can define the $c$-function \cite{Lob}
\be
c_D\simeq \frac{1}{\left( G_{\mu \nu }n^{\mu }n^{\nu }\right) ^{\frac{D}{2}-1}},
\ee
where $n^{\mu }$ is the unit vector along the time direction. The monotonically increasing behaviour of $c$-function as a function of time indicates the flow of renormalization group is toward UV for any expanding patch of dS spacetime. Moreover, decreasing of $c$-function with time shows that the renormalization flow is toward IR for the contracting patch of dS spacetime. As an example, for six-dimensional solutions given in (\ref{HTN6cos}), the behaviour of $c$-metric in figure \ref{fig:c} shows that the six dimensional asymptotically dS spacetime is an expanding patch of dS. 
The $c$-functions for $D$-dimensional solutions (\ref{HTNDcos}) and (\ref{HTNDsecondcos}) show similar monotonically increasing behaviours versus time, hence the renormalization flows are toward UV in any dimensionality.

\section{Concluding Remarks}
\label{sec:con}

In this article, we construct exact convoluted solutions to the six and higher dimensional Einstein-Maxwell theory. The two different classes of solutions in $D$-dimensions are given by the line element (\ref{mD}), gauge field (\ref{gaugeD}) and the metric functions (\ref{HTN7}) and (\ref{HTNS}). The integral measure functions for any solutions can be determined and fixed by considering the solutions in some appropriate limits and comparing them with the solutions in $D$-dimensions (\ref{nonut}) where the metric function (\ref{H09}) has a simple form. Except on the location of a bolt at the origin of spherical coordinates, all the metric functions are regular in any points of spacetime. The bolt singularity may be converted to a higher dimensional regular hypersurface if we consider the metric functions that depend on more than two spatial directions. 
Moreover, we construct the solutions to Einstein-Maxwell theory with positive cosmological constant where the metric function depends on time and two spatial directions. The solutions are given by the metric (\ref{mD}) and gauge field (\ref{gaugeD}) and the cosmological metric functions (\ref{HTNDcos}) and (\ref{HTNDsecondcos}). All cosmological convoluted solutions have monotonically increasing $c$-functions in agreement with the $c$-theorem for asymptotically dS spacetimes.

\bigskip
{\Large Acknowledgments}

This work was supported by the Natural Sciences and Engineering Research
Council of Canada. The author would like to thank the Mainz Institute for Theoretical Physics (MITP) for its hospitality and its partial support during the completion of this work.

\end{document}